\RequirePackage{fix-cm}
\documentclass[twocolumn,epjc3]{svjour3}
\pdfoutput=1
\smartqed  
\RequirePackage{graphicx}
\RequirePackage{mathptmx}      
\RequirePackage{flushend}
\RequirePackage[colorlinks,citecolor=blue,urlcolor=blue,linkcolor=blue]{hyperref}
\hypersetup{
  pdfauthor={Christian Guetschow, Jonas Linder, Marek Schoenherr},
  pdftitle={Multi-jet merged top-pair production including electroweak corrections}
}

\journalname{Eur. Phys. J. C}

\usepackage{cite}
\usepackage{appendix}
\usepackage{epsfig}
\usepackage{amsmath}
\usepackage{amssymb}
\usepackage{bm}
\usepackage{slashed}
\usepackage[normalem]{ulem} 
\usepackage{subfig}
\usepackage{xspace}

\numberwithin{equation}{section}
\allowdisplaybreaks[2]

\newcommand{\CSS}{{\rmfamily\scshape Csshower}\xspace}
\newcommand{\Comix}{{\rmfamily\scshape Comix}\xspace}
\newcommand{\Amegic}{{\rmfamily\scshape Amegic}\xspace}

\newcommand{\Sherpa}{{\rmfamily\scshape Sherpa}\xspace}
\newcommand{\OpenLoops}{{\rmfamily\scshape OpenLoops}\xspace}

\newcommand{\Rivet}{{\rmfamily\scshape Rivet}\xspace}
\newcommand{\Lhapdf}{{\rmfamily\scshape Lhapdf}\xspace}

\newcommand{\intfer}{\text{sub}\xspace}

\newcommand{\NLO}{\text{NLO}\xspace}

\newcommand{\QCD}{\text{QCD}\xspace}

\newcommand{\EW}{\text{EW}\xspace}

\newcommand{\EWvirt}{\ensuremath{\EW_\text{virt}}\xspace}

\newcommand{\QCDpEW}{\text{\QCD{}+\EW}\xspace}

\newcommand{\QCDpEWvirt}{\text{\QCD{}+\EW{}$_\mathrm{virt}$}\xspace}

\newcommand{\CKKW}{\text{CKKW}\xspace}

\newcommand{\MCatNLO}{M\protect\scalebox{0.8}{C}@N\protect\scalebox{0.8}{LO}\xspace}

\newcommand{\MEPS}{M\scalebox{0.8}{E}P\scalebox{0.8}{S}\xspace}
\newcommand{\MENLOPS}{ME\protect\scalebox{0.8}{NLO}PS\xspace}
\newcommand{\MEPSatLO}{M\protect\scalebox{0.8}{E}P\protect\scalebox{0.8}{S}@L\protect\scalebox{0.8}{O}\xspace}
\newcommand{\MEPSatNLO}{M\protect\scalebox{0.8}{E}P\protect\scalebox{0.8}{S}@N\protect\scalebox{0.8}{LO}\xspace}

\newcommand{\mr}[1]{{\ensuremath{\mathrm{#1}}}}
\newcommand{\mc}[1]{{\ensuremath{\mathcal{#1}}}}
\newcommand{\mb}[1]{{\ensuremath{\mathbf{#1}}}}
\newcommand{\done}{\ensuremath{\mr{d}}}

\newcommand{\alphaS}{\ensuremath{\alpha_\mathrm{s}}}
\newcommand{\order}{{\ensuremath{\mc{O}}}}
\newcommand{\Qcut}{{\ensuremath{Q_{\mathrm{cut}}}}}
\newcommand{\nmax}{{\ensuremath{n_{\mathrm{max}}}}}

\newcommand{\nmaxnlo}{{\ensuremath{n_{\mathrm{max}}^{\text{\NLO}}}}}
\newcommand{\core}{{\ensuremath{\text{core}}}}

\newcommand{\pT}{{\ensuremath{p_\mr{T}}}}
\newcommand{\Gmu}{{\ensuremath{G_\mu}}}

\newcommand{\ttbar}{{\ensuremath{t\bar{t}}}\xspace}
\newcommand{\ttbarj}{{\ensuremath{t\bar{t}+\text{jet}}}\xspace}
\newcommand{\shortequal}{{\ensuremath{\!\!\!=\!\!\!}}}
\newcommand{\muR}{{\ensuremath{\mu_R}}}
\newcommand{\muF}{{\ensuremath{\mu_F}}}
\newcommand{\muQ}{{\ensuremath{\mu_Q}}}
\newcommand{\muCKKW}{{\ensuremath{\mu_\CKKW}}}
\newcommand{\mucore}{{\ensuremath{\mu_\core}}}
\newcommand{\zthr}{{\ensuremath{z_{\mathrm{thr}}}}}

\usepackage{xcolor}
\definecolor{darkblue}{cmyk}{1,0.3,0,0.2}
\definecolor{violet}{cmyk}{0,1,0,0.2}
\hypersetup{colorlinks, bookmarksnumbered, citecolor=darkblue, linkcolor=darkblue, pdfstartview=FitH, urlcolor=darkblue, linktocpage}

\begin{document}

\title{Multi-jet merged top-pair production including electroweak corrections}

\author{Christian G\"utschow\thanksref{ucl}, Jonas M. Lindert\thanksref{ippp}, Marek Sch\"onherr\thanksref{cern}}

\institute{%
  Department of Physics and Astronomy, University College London, Gower Street, London, WC1E 6BT, UK \label{ucl}
  \and
  Institute for Particle Physics Phenomenology, Department of Physics, Durham University, Durham, DH1 3LE, UK \label{ippp}
  \and
  Theoretical Physics Department, CERN, CH-1211 Geneva 23, Switzerland\label{cern}
}

\maketitle

\begin{picture}(0,0)
  \put(285,180){CERN--TH--18--039, IPPP/18/13, MCnet-18-05}
\end{picture}

\begin{abstract} 
  We present theoretical predictions for the production of top-quark pairs 
  in association with jets at the LHC including electroweak (EW) 
  corrections. 
  First, we present and compare differential predictions at the fixed-order 
  level for \ttbar and  \ttbarj production at the LHC considering 
  the dominant NLO EW corrections of order $\order(\alphaS^2 \alpha)$ and 
  $\order(\alphaS^3 \alpha)$ respectively together with all additional 
  sub\-leading Born and one-loop con\-tri\-bu\-tions. 
  The NLO EW corrections are enhanced at large energies and in particular 
  alter the shape of the top transverse momentum distribution, whose 
  reliable modelling is crucial for many searches for new physics at the 
  energy frontier. 
  Based on the fixed-order results we motivate an approximation of the EW 
  corrections valid at the percent level, that allows us to readily 
  incorporate the EW corrections in the \MEPSatNLO framework of \Sherpa
  combined with \OpenLoops. 
  Subsequently, we present multi-jet merged parton-level predictions for 
  inclusive top-pair production incorporating NLO QCD+EW corrections to \ttbar 
  and \ttbarj.
  Finally, we compare at the particle-level 
  against a recent 8~TeV measurement of the top transverse momentum 
  distribution performed by ATLAS in the lepton+jet channel. 
  We find very good agreement between the Monte Carlo prediction and the 
  data when the EW corrections are included.
\end{abstract}


\section{Introduction}
\label{sec:intro}

The study of top-quark production and decay plays a key role in the ongoing
physics programme of the LHC. Measurements in the different production modes can
be used for detailed exploration of top-quark interactions as well as properties
such as the top-quark mass, which is one of the fundamental parameters of the
Standard Model (SM). At the same time, top-quark production and in particular
top-quark pair production represents an important and challenging background in
many searches for physics beyond the Standard Model (BSM). The sensitivity of
many of these searches depends in a critical way on the precision of theoretical
simulations in particular in the tails of kinematic distributions.

Theoretical predictions for \ttbar production (in associateion with jets) at ha\-dron colliders are
indeed very advanced. For on-shell top-quarks higher-order corrections have been
calculated fully differentially up to NNLO in the strong
coupling~\cite{Czakon:2013goa,Czakon:2015owf} and up to NLO in the 
EW coupling~\cite{Kuhn:2006vh,Bernreuther:2006vg,Kuhn:2013zoa,Bernreuther:2010ny,
Hollik:2011ps,Pagani:2016caq}. These calculations have also been combined within
a joint setup~\cite{Czakon:2017wor,Czakon:2017lgo}. In fact, here also the
subleading one-loop corrections, first considered in~\cite{Pagani:2016caq}, have been included. 
Considering \ttbar production in association with additional jets, for
\ttbarj~\cite{Dittmaier:2007wz,Melnikov:2010iu}, $t\bar
t+2$ jets~\cite{Bredenstein:2009aj,Bredenstein:2010rs,Bevilacqua:2009zn,Bevilacqua:2010ve,Bevilacqua:2011aa}
and even $t\bar t+3$ jets~\cite{Hoche:2016elu} higher-order corrections are known at NLO QCD reducing
the otherwise significant theoretical uncertainties in the modelling of top-pair plus multi-jet signatures, relevant for
$t\bar t H$ and new-physics searches.
Beyond the on-shell approximation, NLO QCD corrections for combined \ttbar production and decay
have first been studied in the narrow\--width\--approxi\-ma\-tion (NWA)~\cite{Melnikov:2009dn,Bernreuther:2010ny,Campbell:2012uf}
and later also fully off-shell based on the $W^+W^-b\bar b$ final-state with off-shell leptonic~\cite{Bevilacqua:2010qb,Denner:2010jp,Denner:2012yc,Cascioli:2013wga,Frederix:2013gra} and semi-leptonic~\cite{Denner:2017kzu} decays. Including off-shell leptoninc decays also the NLO EW corrections are known~\cite{Denner:2016jyo}.
Furthermore, NLO QCD corrections to top-pair production in association with an
additional jet combining corrections in production and decay have been computed
in the NWA \cite{Melnikov:2011qx} and also fully off-shell with leptonic
decays~\cite{Bevilacqua:2015qha}. Finally, in
\cite{Pagani:2016caq,Czakon:2017wor} the photon-initiated production of
top-quark pairs has been studied.

Particle-level Monte Carlo generators matching \linebreak NLO~QCD matrix elements to parton
showers have been available for quite some time for inclusive \ttbar
production~\cite{Frixione:2003ei,Frixione:2007nw} and since recently for
\ttbarj~\cite{Kardos:2011qa,Alioli:2011as,Czakon:2015cla} and $t\bar
t+2$~jet~\cite{Kardos:2013vxa,Cascioli:2013era,Hoeche:2013mua,Hoeche:2014qda,Bevilacqua:2017cru,Jezo:2018yaf}
production. In fact, in \cite{Hoeche:2014qda} a unified description of top-pair plus multi-jet production has been presented merging $t\bar t+$0,1,2 jets at NLO QCD within the \MEPSatNLO framework of {\sc Sherpa+OpenLoops}~\cite{Hoeche:2009rj,Hoeche:2012yf,Gehrmann:2012yg}. The NLO QCD computation for combined
top-pair production and decay in the NWA has been matched to parton showers \cite{Campbell:2014kua}, and the matching of fully off-shell NLO QCD
top-pair production including leptonic decays was presented in~\cite{Jezo:2016ujg}. The latter required the
 development of a modified resonance-aware NLO matching scheme~\cite{Jezo:2015aia}.

Experimental cross-section measurements for top-pair production at the LHC are
similarly advanced. After initial measurement at the inclusive cross section
level~\cite{Chatrchyan:2012bra,Chatrchyan:2013faa,Aad:2014kva,Khachatryan:2015uqb,Aaboud:2016pbd,Nayak:2017aes},
where very good agreement with perturbative calculations at the NNLO+NNLL level
in QCD~\cite{Cacciari:2011hy,Beneke:2011mq,Czakon:2013goa} has been observed,
the attention in the study of top-pair production has shifted towards detailed
differental measurements, see
e.g.~\cite{Chatrchyan:2012saa,Aad:2014zka,Aad:2015eia,Khachatryan:2015oqa,Aad:2015hna,
Khachatryan:2016mnb,Sirunyan:2017mzl,Aaboud:2016syx,Aaboud:2018eqg}. One of the
most important observables in \ttbar production, in particular relevant for
beyond the Standard Model searches at the energy frontier, is the transverse
momentum distribution of (reconstructed) top quarks. Different measurements of
this observables consistently indicate that the top quark transverse momentum
distribution at low $\pT$ is well predicted by the employed Monte Carlo
programs, both in normalisation and shape, but these predictions exceed the data
at high $\pT$. Comparing these measurements at the unfolded parton level to
differential NNLO QCD predictions~\cite{Czakon:2015owf}, this excess has been
alleviated. This indicates the relevance of including higher jet multiplicities
for the modelling of the top transverse momentum spectrum at large $\pT$. At the
same time at large $\pT$ the higher-order EW corrections are enhanced due to the
appearance of EW Sudakov
logarithms~\cite{Kuhn:2006vh,Bernreuther:2006vg,Kuhn:2013zoa,Bernreuther:2010ny,
Hollik:2011ps,Pagani:2016caq,Denner:2016jyo} yielding shape distortions at the
level of $-10\%$ for $p_\text{T,top}=1$~TeV.

In this paper we present multi-jet merged predictions for top-pair production
including QCD and EW corrections. To this end, we first present the original
calculation of \ttbarj production at NLO EW, i.e.\ the corrections of
$\order(\alphaS^3 \alpha)$, and we also consider the subleading one-loop
corrections of $\order(\alphaS^2 \alpha^2)$ and $\order(\alphaS \alpha^3)$. We
compare these corrections with the corresponding ones for inclusive \ttbar
production. This comparison allows to estimate non-factorising NNLO mixed QCD-EW
contributions to \ttbar production. Furthermore, we present predictions merging
\ttbar and \ttbarj production at NLO QCD+EW within the \MEPSatNLO multi-jet
merging framework in {\sc Sherpa} combined wit \OpenLoops, incorporating the EW
corrections in an approximation~\cite{Kallweit:2015dum} that we show holds at
the one percent level. In this approximation, the dominant virtual NLO EW
corrections are incorporated exactly, while the NLO QED bremsstrahlung is first
integrated out and subsequently incorporated via YFS multi-photon
emission~\cite{Schonherr:2008av}. Finally, we compare the resulting \MEPSatNLO
QCD+EW$_{\rm virt}$ predictions, including spin-correlation preserving top-quark
decays at LO, for the reconstructed top-quark transverse momentum distribution
at particle-level against a recent measurement performed by ATLAS in the
lepton+jet channel at 8 TeV, based on a selection of top-quark candidates in the
boosted regime~\cite{Aad:2015hna}. We find very good agreement between the Monte
Carlo prediction and the data when the EW corrections are included.

The structure of this paper is as follows. In Section \ref{sec:fo} we present
fixed-order predictions for $\ttbar$ and $\ttbarj$ production including all
one-loop electroweak corrections. In Section \ref{sec:meps} we present the
methodology of incorporating the NLO EW corrections in the \MEPSatNLO framework.
Resulting predictions at parton- and particle-level merging the zero and one jet
multiplicties at NLO QCD+EW are presented in Section \ref{sec:results}. We
conclude in Section \ref{sec:conclusions}.

\section{Electroweak corrections for \texorpdfstring{$\boldsymbol{pp\to t\bar{t}+0,1}$ jet}{pp->tt+0,1jet}}
\label{sec:fo}

In the following we present electroweak corrections to the processes
\begin{equation}
pp \to \ttbar \quad \text{and} \quad pp \to \ttbar+{\rm jet} \,,
\end{equation}
which are described at leading order (LO) at $\order(\alphaS^2)$ 
and $\order(\alphaS^3)$, respectively.
Additionally, there are subleading Born contributions of 
$\order(\alphaS \alpha)$ and $\order(\alpha^2)$ to 
$pp \to \ttbar$ production and of $\order(\alphaS^2 \alpha)$ and 
$\order(\alphaS \alpha^2)$ to $pp \to \ttbarj$ production.\footnote{
  In our calculation we do not consider photon induced processes. 
  As shown in \cite{Czakon:2017wor} these contribute only at the one percent level 
  for \ttbar production given PDFs based on the LUXqed methodology~\cite{Manohar:2017eqh} are used. 
  We verified that this also holds for \ttbarj production, where relative contributions 
  of photon-induced production are even smaller compared to \ttbar production. 
} 
The $\order(\alphaS \alpha)$ contribution to \ttbar production 
is strongly suppressed as only $b$-quark-initiated processes contribute where 
diagrams involving a $t$-channel $W^{\pm}$ exchange interfere with 
diagrams involving an $s$-channel gluon.
All other contributions vanish due to their colour structure. 
In \ttbarj production all $q\bar q$ channels contribute at both 
subleading Born orders. 

At the one-loop level the electroweak corrections comprise $\order(\alphaS^2
\alpha)$, $\order(\alphaS \alpha^2)$ and $\order(\alpha^3)$ contributions to
\ttbar production, and $\order(\alphaS^3 \alpha)$, $\order(\alphaS^2 \alpha^2)$
and $\order(\alphaS \alpha^3)$ contributions to \ttbarj production. Customarily,
the leading one-loop electroweak contributions are denoted as NLO EW
corrections, i.e.\ the relative corrections of $\order(\alpha)$ with respect to
the LO processes. At large energies these contributions develop a logarithmic
en\-hance\-ment that factorises from the respective LO
contributions~\cite{Denner:2000jv,Denner:2001gw,Denner:2008yn}.

In order to introduce the notation, we define the components of 
a $\order(\alphaS^i \alpha^j)$ Born-level computation as
\begin{equation}\label{eq:lo}
  \done\sigma^\text{LO}_{ij} =\; \done\Phi_B\, \mr{B}_{ij}(\Phi_B)
\end{equation}
Therein, $\mr{B}_{ij}$ is the $\order(\alphaS^i \alpha^j)$  matrix element including 
all PDF and symmetry/averaging factors and $\done\Phi_B$ is 
its accompanying phase-space configuration. As a shorthand notation corresponding predictions 
are denoted as LO$_{ij}$. 
Correspondingly, we define a one-loop correction of $\order(\alphaS^i \alpha^j)$ as
\begin{equation}\label{eq:nlo}
  \begin{split}
    \done\sigma^{\Delta\text{NLO}}_{ij}
    =\;&\done\Phi_B\,\tilde{\mr{V}}_{ij}(\Phi_B)
        +\done\Phi_R\,\mr{R}_{ij}(\Phi_R)\;.
  \end{split}
\end{equation}
Here, $\mr{R_{ij}}$ and $\done\Phi_R$ denote the corresponding real-emission
matrix element and phase space, respectively, while 
$\tilde{\mr{V}}_{ij}$ contains the virtual correction 
$\mr{V}_{ij}$ as well as the collinear counterterm 
of the PDF mass factorisation. Again as a short-hand notation these
one-loop corrections are denoted as $\Delta$NLO$_{ij}$.

All these contributions to $\ttbar$ and $\ttbarj$ production can readily be 
computed within the {\sc Sherpa+OpenLoops} framework.\footnote{
  The extension of these tools to provide higher-order electroweak corrections 
  will very soon be publicly released.
} 
In this framework the virtual corrections are computed with the \OpenLoops
amplitude provider~\cite{Cascioli:2011va,OLhepforge,Kallweit:2014xda}, which
implements a very fast hybrid tree-loop recursion to construct and compute
one-loop scattering amplitudes in the full SM. For the integral reduction
\OpenLoops is interfaced with {\sc Collier}~\cite{Denner:2016kdg} and {\sc
CutTools}~\cite{Ossola:2007ax}.
The tree-level matrix elements as well as the infrared subtraction, process
management and phase-space integration of all contributing partonic channels,
are provided by \Sherpa through its tree-level matrix element generator \Amegic
\cite{Krauss:2001iv}. In \linebreak \Sherpa, infrared divergences are subtracted
using a generalisation of the Catani-Seymour scheme \cite{Catani:1996vz,
Catani:2002hc,Gleisberg:2007md,Dittmaier:2008md,Schonherr:2017qcj}, used
previously in \cite{Kallweit:2014xda,Kallweit:2015dum,Biedermann:2017yoi,
Kallweit:2017khh,Chiesa:2017gqx,Greiner:2017mft}, and include the appropriate
initial state mass factorisation counter terms. Cross-checks of the renormalised
pole coefficients of the virtual corrections computed by \OpenLoops and the
infrared poles supplied by \Sherpa have been performed and excellent agreement
has been found.

For our predictions of \ttbar and \ttbarj production we choose input parameters in 
accordance with Table \ref{tab:inputs}. The electroweak coupling $\alpha$ is fixed and
renormalised according to the $G_\mu$-scheme, $\alpha = \frac{\sqrt{2}}{\pi}\,
G_{\mu} \left| \mu_{W}^2 \sin_{\theta_{\rm w}}^2 \right|$, guaranteeing an optimal 
description of pure SU(2) interactions at the \EW scale.
Here, $\mu_{W}$ denotes the complex-valued $W$ mass, with $\mu_V^2=m_V^2-i\Gamma_V
m_V$ and $\theta_{\rm w}$ the equally complex valued weak mixing angle, derived
from the ratio $\mu_W/\mu_Z$. The massive vector bosons and the Higgs are
renormalised in the complex-mass scheme \cite{Denner:2005fg}, while the top-quark is kept stable and
correspondingly renormalised in the on-shell scheme. The introduction of finite
widths for the massive vector bosons is mandatory due to the appearance of
otherwise singular resonant internal propagators in the 
bremsstrahlung to \ttbarj production.
As renormalisation and factorisation scales for the strong
coupling $\alphaS$ we use 
\begin{equation}
\label{eq:foscale}
\mu=\mu_R=\mu_F=\frac{1}{2}(E_{\text{T},t} + E_{\text{T},\bar
t})\;, 
\end{equation}
where $E_{\text{T},t/\bar{t}}$ denotes the transverse energy of the top/antitop.
In the predictions for top-pair plus jet production we recombine collinear
photon--quark pairs within a cone of $R_{\gamma q} < 0.1$ and 
cluster jets according to the anti-$k_{\rm T}$ algorithm implemented in {\sc
FastJet}~\cite{Cacciari:2011ma} and require
\begin{equation}
  p_{T,j} > 30~\text{GeV}\;, \quad \text{and} \quad |\eta_{j}| < 4.5 \;.
\end{equation}
Jets with a photonic energy fraction larger than $\zthr=0.5$ 
are discarded. 
No phase space cuts are applied to the final state top quarks.

\begin{table}[t]
  \begin{center}
    \begin{tabular}{rclrcl}
      $\Gmu$ & \shortequal & $1.1663787\cdot 10^{-5}~\text{GeV}^2$ 
        & \qquad & &\\
      $m_W$ & \shortequal & $80.385~\text{GeV}$  
        & $\Gamma_W$ & \shortequal & $2.0897~\text{GeV}$ \\
      $m_Z$ & \shortequal & $91.1876~\text{GeV}$ 
        & $\Gamma_Z$ & \shortequal & $2.4955~\text{GeV}$ \\
      $m_h$ & \shortequal & $125~\text{GeV}$     
        & $\Gamma_h$ & \shortequal & $4.07~\text{MeV}$ \\
      $m_t$ & \shortequal & $173.2~\text{GeV}$   
        & $\Gamma_t$ & \shortequal & $0$ 
    \end{tabular}
  \end{center}\vspace*{-5mm}
  \caption{
    Numerical values of input parameters. While the masses are 
    taken from \cite{Agashe:2014kda}, the widths are obtained from 
    state-of-the-art calculations.
    \label{tab:inputs}
  }
\end{table}

We work in the five-flavour scheme and use the  \linebreak
NNPDF3.0nnlo PDF set \cite{Ball:2014uwa} with $\alphaS=0.118$ 
interfaced through \Lhapdf{}6 \cite{Buckley:2014ana}. 
The initial state QED evolution we thereby neglect has been shown 
to be negligible for \ttbar production \cite{Pagani:2016caq}.
Here we present results for the LHC at a centre-of-mass energy of 13\,TeV. 
However, the relative
electroweak corrections are fairly insensitive to the centre-of-mass energy,
and choices of $\mu_R$ and $\mu_F$.

\begin{figure*}[t]
  \includegraphics[width=.48\textwidth]{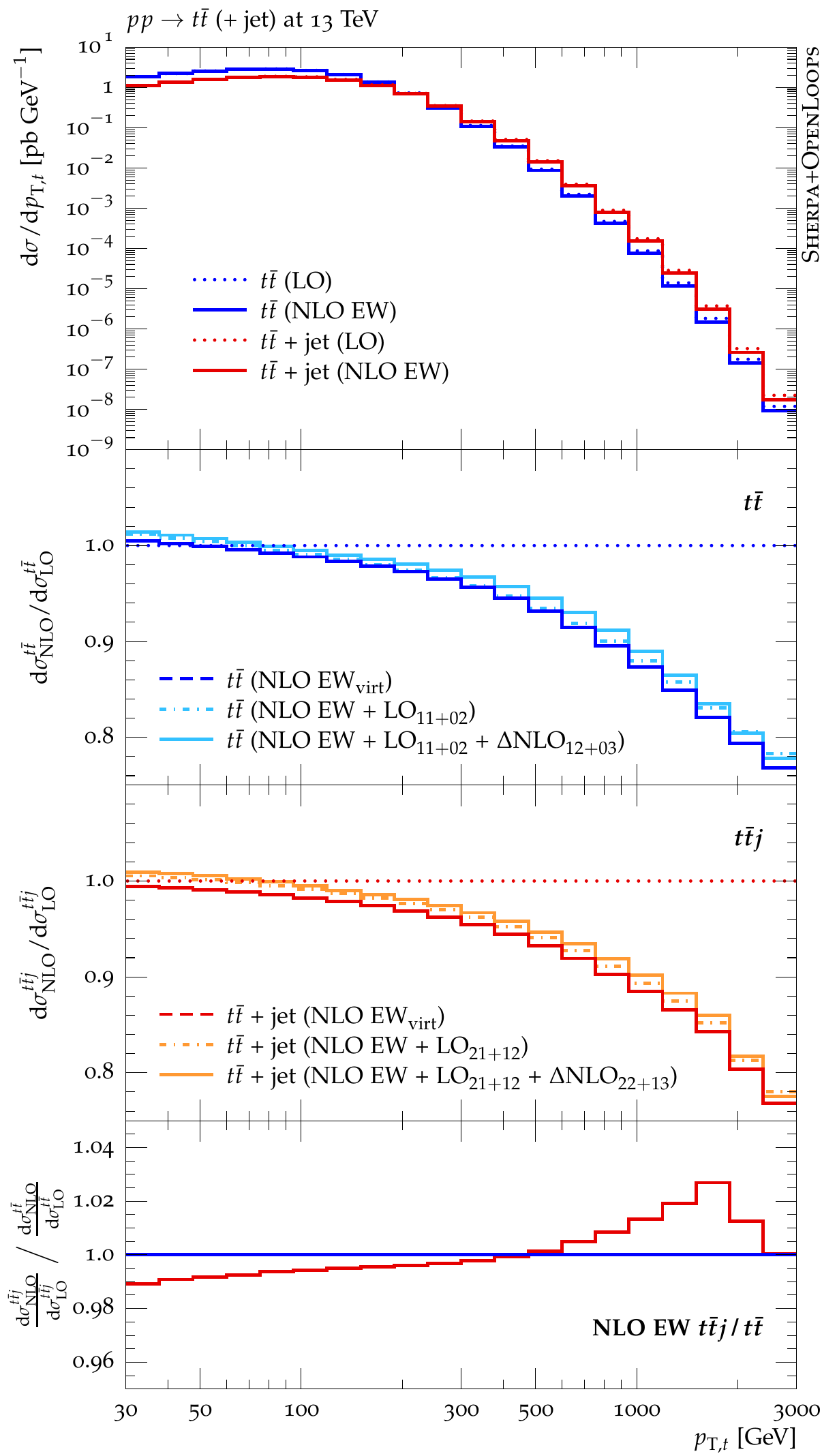}
  \hfill
  \includegraphics[width=.48\textwidth]{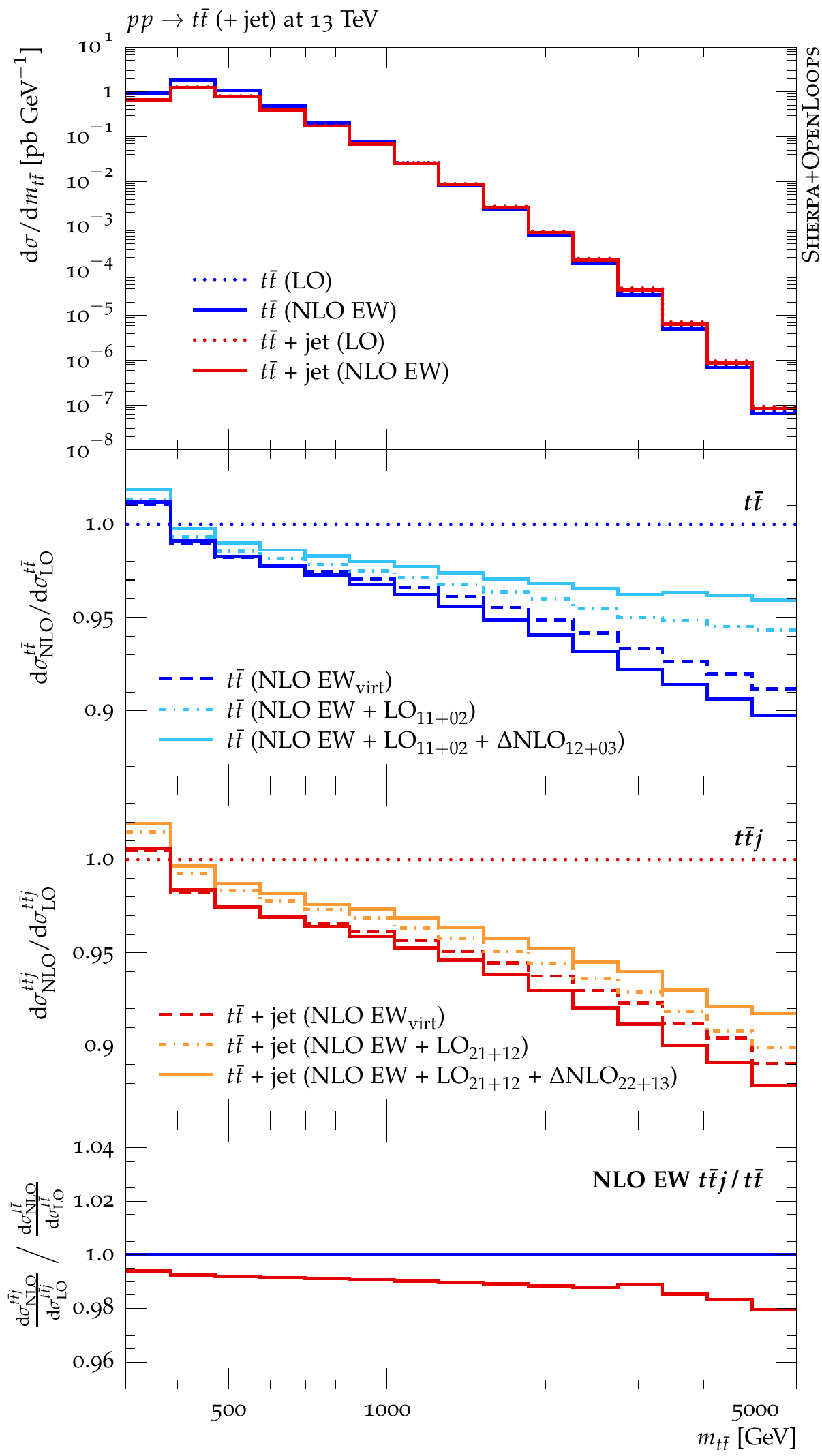}
  \caption{
    Top quark transverse momentum (left) and top-antitop 
    invariant mass (right) in inclusive $\ttbar$ production 
    (blue) and $\ttbarj$ production (red) at 
    NLO EW at 13\,TeV at the LHC. In $\ttbarj$ we require $\pT>30\,\text{GeV}$.
    The top panel shows the differential cross section, 
    while the three lower panels show, from top to bottom, 
    the subleading Born and higher-order corrections to inclusive $\ttbar$ production 
    and $\ttbarj$ production, respectively. Subleading Born and one-loop contributions are shown with 
    lighter shades of the colour of the respective processes, dashed lines 
    containing only the subleading Born contributions and 
    solid lines containing all subleading Born and one-loop contributions. 
    The lowest panel shows the ratio of the NLO EW corrections to the two processes.
    Corrections based on the NLO EW$_{\rm virt}$ approximation are shown 
    as the dashed line of the same colour as the exact NLO EW result.
    \label{fig:pTtop_mtt}
  }
\end{figure*}

In Figure \ref{fig:pTtop_mtt} we show the NLO \EW corrections together with the
effect of the subleading Born and one-loop contributions to the top quark
transverse momentum and the $\ttbar$ invariant mass in $\ttbar$ and $\ttbarj$
production. We find that the NLO EW corrections exhibit the expected electroweak
Su\-da\-kov-like shape. They are small at low transverse momenta and invariant
masses and continuously grow to reach \linebreak $-10(20)\%$ at a top-quark
transverse momentum of 1(2)\,TeV and $-5(10)\%$ at $\ttbar$ invariant masses of
2(4)\,TeV, respectively. The NLO EW corrections to the $\ttbarj$ process
reproduce those to the inclusive $\ttbar$ process to very high accuracy in these
observables. Their ratio never exceeding a few percent. Thus, the electroweak
corrections factorise to a very good approximation with respect to additional
jet activity. This finding supports a multiplicative combination of QCD and NLO
EW corrections in \ttbar production.

In Figure \ref{fig:pTtop_mtt} subleading Born and one-loop contributions are
shown with lighter shades of the colour of the respective processes, dash-dotted
lines containing only the subleading Born contributions and solid lines
containing all subleading Born and one-loop contributions. The subleading Born
and one-loop orders contribute only at the percent level to the transverse
momentum distribution, both for $\ttbar$ and $\ttbarj$ production. In
particular, while for both $\ttbar$ and $\ttbarj$ the subleading Born
contributions (LO$_{11}$ and LO$_{21}$, respectively) are negligible, the
inclusion of the sub-subleading Born contributions (LO$_{02}$ and LO$_{12}$,
respectively) increase the cross section an almost constant $\approx 1\%$. On
the other hand, for the invariant mass of the $\ttbar$-pair these subleading
Born contributions show a different behaviour for inclusive $\ttbar$ production
and $\ttbar$ production in association with a jet. While for the latter we again
observe an almost constant increase of the cross section by 1-2\%, there is a
clear shape distortion induced in the case of inclusive $\ttbar$ production.
Here, for $m_{\ttbar}>2$\,TeV the subleading Born contributions, largely
dominated by the LO$_{21}$ contribution, amount to $\approx\tfrac{1}{2}$ the NLO
EW correction, only with opposite sign. The resulting compensation needs to be
accounted for. Here, the subleading one-loop corrections are dominated by the
$\Delta$NLO$_{22}$ contributions and can in some sense be understood as the NLO
QCD corrections to the sub-subleading Born of LO$_{12}$. However, we want to
note that the $\order(\alphaS^2 \alpha^2)$ bremsstrahlung also comprises $ttV$
production with $V\to q \bar q$ decays, where $V=\{W^{\pm},Z\}$. Thus, in
principle care has to be taken when such processes are considered as separate
backgrounds in BSM searches. However, these subleading one-loop corrections
contribute only at the percent level, with an increasing effect at very large
$m_{\ttbar}$.

\begin{figure}[t!]
  \includegraphics[width=.48\textwidth]{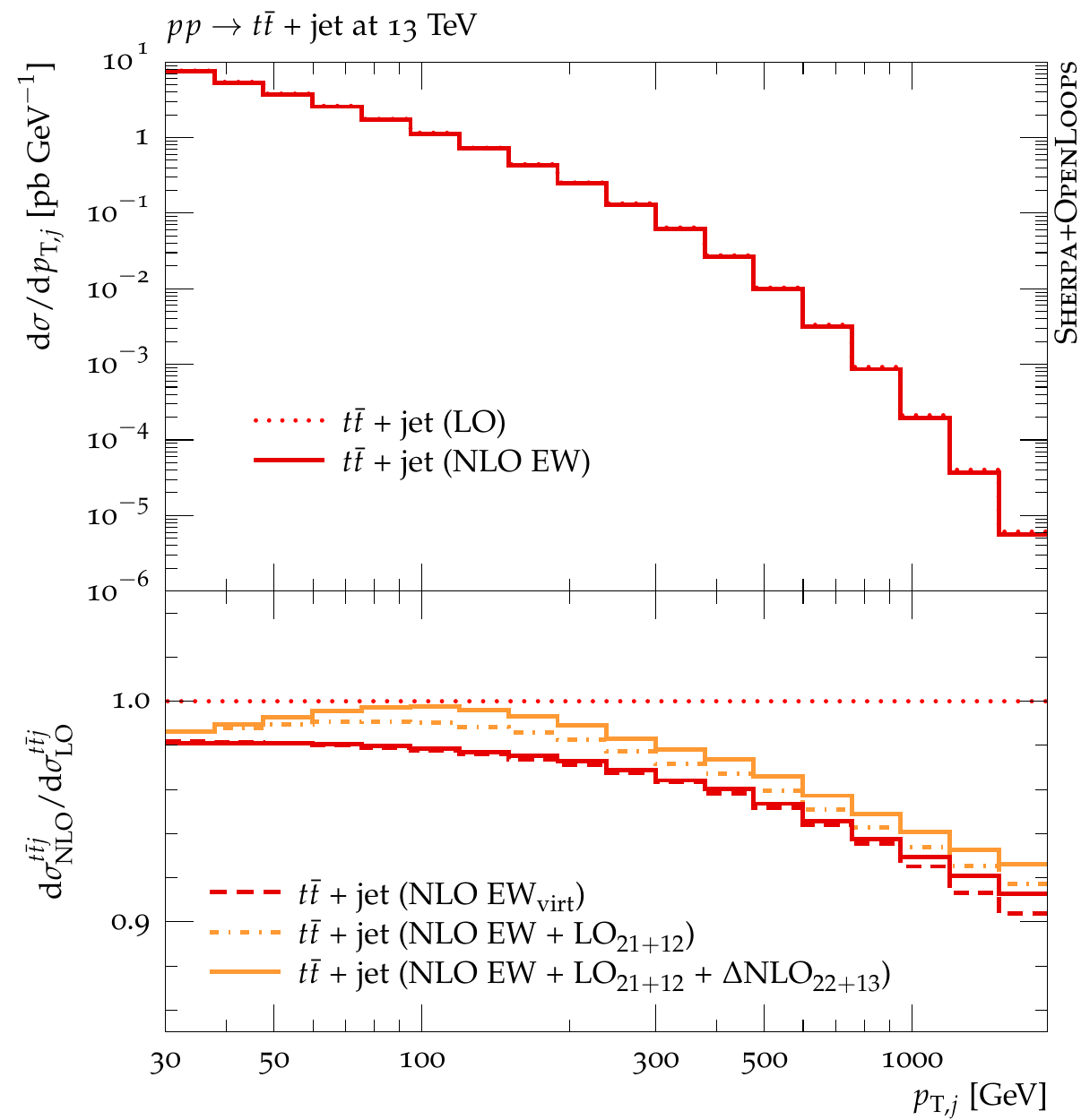}
  \caption{
    Leading jet transverse momentum in $\ttbarj$ production at 
    NLO EW at 13\,TeV at the LHC. 
    The top panel shows the differential cross section, 
    while the lower panel shows the subleading Born and one-loop electroweak corrections.
    Colour and line coding as in Figure \ref{fig:pTtop_mtt}. 
  }
\label{fig:pTjet1}
\end{figure}

In Figure  \ref{fig:pTtop_mtt} we also investigate the quality of the so-called 
$\EWvirt$ approximation \cite{Kallweit:2015dum} 
defined as
\begin{equation}\label{eq:ewvirt}
  \begin{split}
    \done\sigma^\text{NLO \EWvirt}
    =\;&\done\Phi_B\,
	  \left[
	    \mr{B}_{(n+2)0}(\Phi_B)
	    +\mr{V}_{(n+2)1}(\Phi_B)
	    \vphantom{\int_1}
	  \right.\\
       &\left.\hspace*{10mm}{}
	    +\int_1\done\Phi_1\,\mr{R}_{(n+2)1}^\text{approx}(\Phi_B\cdot\Phi_1)
	  \right]\;,\hspace*{-5mm}
  \end{split}
\end{equation}
where $n$ denotes the jet multiplicity in  $\ttbar+n~$jet production.
We define the approximated real-emission contribution 
$\mr{R}_{(n+2)1}^\text{approx}$ such that its integral over the 
real-emission phase space equals the standard Catani-Seymour 
$\mb{I}$-operator. 
This approximation is both finite and 
local in the Born phase space and can hence be easily applied 
as a corrective weight in the multijet merging introduced in  
Section \ref{sec:meps}. 
By construction, it is expected to correctly reproduce the 
exact NLO EW corrections in the Sudakov limit, but also 
contain important non-logarithmic terms extending its 
validity in practice.

In Figure \ref{fig:pTtop_mtt} the result 
using this approximation is detailed as the dashed line 
of the same colour as the exact NLO EW result. 
We find generally very good agreement, especially for 
the transverse momenta of the top quark in both $\ttbar$ 
and $\ttbarj$ production. 
Small differences are found for the invariant mass of the 
$\ttbar$-pair at values larger than 1\,TeV, growing to 
relative difference of 1\% at 5\,TeV.

Finally, Figure \ref{fig:pTjet1} details the higher-order EW corrections 
to the leading jet transverse momentum in $\ttbarj$ 
production. 
Similar but smaller corrections as in the top transverse momentum 
distribution are observed. In particular, we find NLO EW corrections of 
about $-10\%$ at 2\,TeV. Subleading Born and one-loop corrections are 
marginally relevant, i.e. they contribute below one percent.
Again, we observe a very good agreement
of the NLO EW$_{\rm virt}$ approximation with the exact NLO EW result.

\section{\texorpdfstring{\MEPSatNLO \QCDpEWvirt}{MEPS@NLO QCD+EWvirt}}
\label{sec:meps}

To illustrate how electroweak corrections can easily be embedded 
in the multijet merging techniques used in the \linebreak\Sherpa Monte Carlo 
event generator, we start with a 
sche\-matic review of the \MEPS method \cite{Catani:2001cc,
  Lonnblad:2001iq,Krauss:2002up,Hoeche:2009rj,Hoche:2010kg,
  Hoeche:2012yf,Gehrmann:2012yg}. 
The aim is to generate inclusive event samples with a variable jet 
multiplicity, wherein the hardest $n=0,1,\ldots,\nmax$ jets, 
according to a measure $Q_n$, are described by the respective 
$n$-jet matrix elements at LO or NLO accuracy. 
A resolution criterion, $\Qcut$, is introduced to separate the 
$n$-jet state from the $n+1$-jet state. 
Thus, in every matrix element $Q_1>\ldots>Q_n>\Qcut$ holds.

In the leading order formulation of this merging method 
\cite{Hoeche:2009rj}, \MEPSatLO, the exclusive cross section 
with exactly $n<\nmax$ jets reads 
\begin{equation}\label{eq:mepsatlo}
  \begin{split}
    \lefteqn{\hspace*{-4mm}\done\sigma_{n}^\text{\MEPSatLO}}\\
    =\;&\done\Phi_{n}\,
          \mr{B}_{n}(\Phi_n)\,\Theta(Q_{n}-\Qcut)\,
          \mc{F}_{n}(\mu_Q^2\,;<\!\Qcut)\;.
  \end{split}
\end{equation}
As in Section \ref{sec:fo}, $\mr{B}_n$ is the relevant Born matrix element including 
all PDF and symmetry/averaging factors, 
and $\Phi_n$ is the $n$-jet phase space configuration. 
Throughout, only the LO contribution, 
$\order(\alphaS^{2+n})$ for $\ttbar+n\,$jet production, 
are considered. 
The $\Theta$-function ensures that all jets are resolved. 
Finally, the parton shower generating functional 
$\mc{F}_{n}(\mu_Q^2\,;<\!\Qcut)$ applies a truncated vetoed 
parton shower to the $n$-jet configuration, ensuring that 
all emissions fall into the unresolved region, $Q<\Qcut$. 
For the highest mulitplicity process, $n=\nmax$, the veto 
is relaxed to $Q_\nmax$ to arrive at a fully inclusive 
description. 
Through its veto it also applies a Sudakov form factor 
weight to the $n$-jet configuration, resumming the 
hierarchy of reconstructed parton shower branchings 
$\mu_Q^2=t_0,t_1,\ldots,t_n$.
Together with the CKKW scale choice, $\muR=\muCKKW$, 
defined through \cite{Catani:2001cc,Bothmann:2016nao} 
\begin{equation}\label{eq:ckkwscale}
  \begin{split}
    \alphaS^{2+n}(\mu_\CKKW^2)
    =\alphaS^2(\mu_\core^2)\cdot\alphaS(t_1)\cdots\alphaS(t_n)\,,
  \end{split}
\end{equation}
and $\muF=\muQ=\mucore$ a smooth transition across $\Qcut$ is 
ensured. 
The core scale is chosen as \cite{Hoeche:2013mua,Hoeche:2014qda} 
\begin{equation}\label{eq:corescale}
  \begin{split}
    \mu_\core
    =\frac{1}{2} \left(\frac{1}{\hat s}+\frac{1}{m_t^2-\hat t}+\frac{1}{m_t^2-\hat u} \right)^{-\frac{1}{2}}.
  \end{split}
\end{equation}
on the reconstructed core $2\to2$ process.

When upgrading the theoretical accuracy of the input $n$-jet 
cross section to NLO QCD accuracy one arrives at the \MEPSatNLO 
method \cite{Hoeche:2012yf,Gehrmann:2012yg}.
Its exclusive $n$-jet cross sections, with $n<\nmaxnlo$, are defined as 
\begin{equation}\label{eq:mepsatnlo}
  \begin{split}\\
    \lefteqn{\hspace*{-4mm}\done\sigma_{n}^\text{\MEPSatNLO}}\\
    \;=&\;\biggl[\done\Phi_{n}\,
          \bar{\mr{B}}_{n}(\Phi_n)\,
          \bar{\mc{F}}_{n}(\mu_Q^2\,;<\!\Qcut)
    \\
    &{}\;\;
    +\done\Phi_{n+1}\,
         \mr{H}_{n}(\Phi_{n+1})\,
         \Theta(\Qcut-Q_{n+1})\,
         \mc{F}_{n+1}(\mu_Q^2\,;<\!\Qcut)\biggr]
    \\
    &{}\times\,\Theta(Q_{n}-\Qcut)\;,
  \end{split}
\end{equation}
based on the \MCatNLO expressions of \cite{Hoeche:2011fd,
  Hoeche:2012ft,Hoche:2012wh}.
Therein, the $\bar{\mr{B}}_{n}$ term 
are the so-called \MCatNLO standard events 
describing the production of $n$ resolved partons 
with $Q_{n}>\Qcut$ at matrix-element level including 
virtual corrections. 
The $(n+1)$-th emission is generated through a fully 
colour- and spin-correlated one-step parton shower 
$\bar{\mc{F}}_{n}$ \cite{Hoeche:2011fd}, restricted 
to the unresolved region. 
The $\bar{\mr{B}}_{n}$ function takes the form 
\begin{equation}\label{eq:mepsatnlo_bbar}
  \begin{split}
    \bar{\mr{B}}_{n}(\Phi_{n})=&\;
      \mr{B}_{n}(\Phi_{n})+\tilde{\mr{V}}_{n}(\Phi_{n})\\
      &{}
      +\int\done\Phi_1\,\mr{D}_{n}(\Phi_{n},\Phi_1)\,\Theta(\mu_Q^2-t_{n+1})\;.
  \end{split}
\end{equation}
wherein $\tilde{\mr{V}}_{n}(\Phi_{n})$ consisting of 
virtual \QCD corrections and initial-state 
collinear mass-factorisation counterterms and 
$\mr{D}_{n}$ is the 
evolution kernel of $\bar{\mc{F}}_{n}$ 
\cite{Hoeche:2012yf} and the $\Theta$-function 
restricts the parton shower phase space. 
The $\mr{H}_{n}$ term, on the other hand, 
corresponds to so-called \MCatNLO hard events. 
Its purpose is to correct the approximate emission 
pattern of the $\bar{\mc{F}}_{n}$ resummation kernels 
and guarantee \NLO \QCD accuracy. 
It is thus also subject to the $(Q_{n+1}<\Qcut)$ 
requirement. 
Again, for $n=\nmaxnlo=\nmax$, the requirement on the 
$(n+1)$-th emission is relaxed to $Q_\nmax$.
The $\mr{H}_{n}$ function takes the form
\begin{equation}\label{eq:mepsatnlo_h}
  \begin{split}
    \mr{H}_{n}(\Phi_{n+1})=&\;
    \mr{R}_{n}(\Phi_{n+1})-
      \mr{D}_{n}(\Phi_{n+1})\,\Theta(\mu_Q^2-t_{n+1})\;,\hspace*{-5mm}
  \end{split}
\end{equation}
wherein $\mr{R}_{n}(\Phi_{n+1})$ denotes the real-emission matrix elements.

It is possible to set $\nmaxnlo<\nmax$, i.e.\ 
describing only the $\nmaxnlo$ lowest jet multiplicities 
at \NLO \QCD accuracy and adding the next $\nmax-\nmaxnlo$ 
jet multiplicities at leading order. 
The resulting \MENLOPS method \cite{Hoche:2010kg,Gehrmann:2012yg,
  Hoeche:2014rya} 
use eq.\ \eqref{eq:mepsatnlo} to describe the multiplicities 
described at \NLO \QCD. 
The subsequent LO multiplicities with $n=\nmaxnlo+k$ ($k>0$) 
are defined through 
\begin{equation}\label{eq:menlops}
  \begin{split}
    \lefteqn{\hspace*{-4mm}\done\sigma_{n}^\text{(\MENLOPS)}}\\
    =\;&\done\Phi_{n}\,
	  k_\nmaxnlo(\Phi_{\nmaxnlo}(\Phi_n))\,
          \mr{B}_{n}(\Phi_n)\,\Theta(Q_{n}-\Qcut)\,\\
       &{}\times\;
          \mc{F}_{n}(\mu_Q^2\,;<\!\Qcut)
  \end{split}
\end{equation}
i.e.\ it supplies the \MEPSatLO expression of eq.\ \eqref{eq:mepsatlo} 
with the local $K$-factor, defined on the highest multiplicity 
described at \NLO \QCD, 
\begin{equation}\label{eq:menlops-kfac}
  \begin{split}
    k_n(\Phi_n)
    \,=\;&
      \frac{\bar{\mr{B}}_n(\Phi_n)}{\mr{B}_n(\Phi_n)}
      \left(
	1-\frac{\mr{H}_n(\Phi_{n+1})}{\mr{B}_{n+1}(\Phi_{n+1})}
      \right)
      +\frac{\mr{H}_{n}(\Phi_{n+1})}{\mr{B}_{n+1}(\Phi_{n+1})}\;.
  \end{split}
\end{equation}
This expression expands to $\order(\alphaS)$ such that the leading order 
accuracy of the $n$-jet sample is not affected. Here \linebreak 
$\Phi_{\nmaxnlo}(\Phi_n)$ denotes the projection of the $n$-jet 
phase space on the $\nmaxnlo$-jet one, taken from the identified 
cluster history constructed during the identification of the 
emission scales $t_i$ \cite{Hoeche:2009rj}. 
This $K$-factor is constructed such that the combination 
of the exclusive $\nmaxnlo$-jet process at \NLO \QCD and 
the inclusive  $(\nmaxnlo+1)$-jet process at LO reproduces 
the inclusive $\nmaxnlo$-jet process at \NLO \QCD exactly 
and, thus, minimises discontinuities across $\Qcut$. 
The inclusion of additional jet multiplicities at LO is in particular 
relevant for the reliable modelling of $t\bar t+$multijet signatures in 
BSM background
simulations.

In order to incorporate approximate \NLO \EW 
corrections in the \MEPSatNLO framework we replace the usual NLO \QCD 
$\bar{\mr{B}}_{n}$ function of eq.\ \eqref{eq:mepsatnlo_bbar} 
with~\cite{Kallweit:2015dum}
\begin{equation}\label{eq:mepsatnloewa}
  \begin{split}
    \lefteqn{\hspace*{-4mm}\bar{\mr{B}}_{n,\QCDpEW}(\Phi_{n})}\\
    =\;&
    \bar{\mr{B}}_{n}(\Phi_{n})
        +{\mr{V}}_{n,\EW}(\Phi_{n})
        +\mr{I}_{n,\EW}(\Phi_{n})
        +\mr{B}_{n,\intfer}(\Phi_{n})\;.\hspace*{-5mm}
  \end{split}
\end{equation}
incorporating NLO EW corrections in the 
\EWvirt approximation defined in eq.\ \eqref{eq:ewvirt}, together with
subleading Born contributions. 
Again, ${\mr{V}}_{n,\EW}(\Phi_{n})$ and $\mr{I}_{n,\EW}(\Phi_{n})$ 
represent the renormalised virtual corrections of 
$\order(\alphaS^{2+n}\alpha)$ and the approximate real emission 
corrections of the same order, integrated over the real emission 
phase space, in the form of the \NLO \EW generalisation of the 
Catani--Seymour $\mathbf{I}$ operator, respectively, as discussed
in Section \ref{sec:fo}.

In the multijet merged setup, the real emission corrections as 
well as the additional LO multiplicities are supplied by 
\Comix \cite{Gleisberg:2008fv} and then showered using the 
\CSS \cite{Schumann:2007mg} based on Catani--Seymour splitting kernels. 
In the following, we perform both a technical validation of 
the presented algorithm at parton-level and then compare 
the full particle-level simulation including spin correlated LO top-quark decays
against data.

\section{Results}
\label{sec:results}

\begin{figure*}[t]
  \includegraphics[width=.48\textwidth]{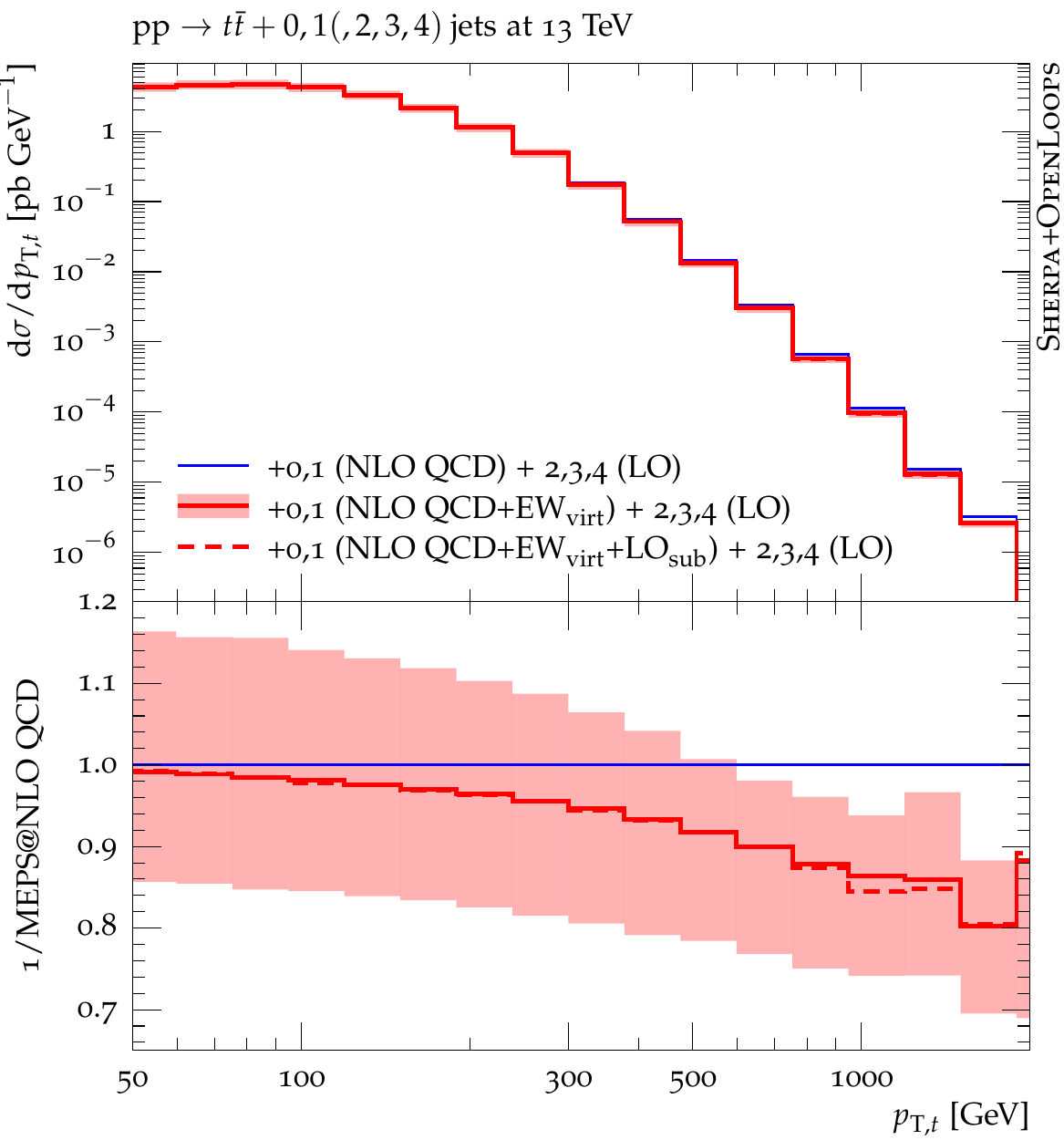}
  \hfill
  \includegraphics[width=.48\textwidth]{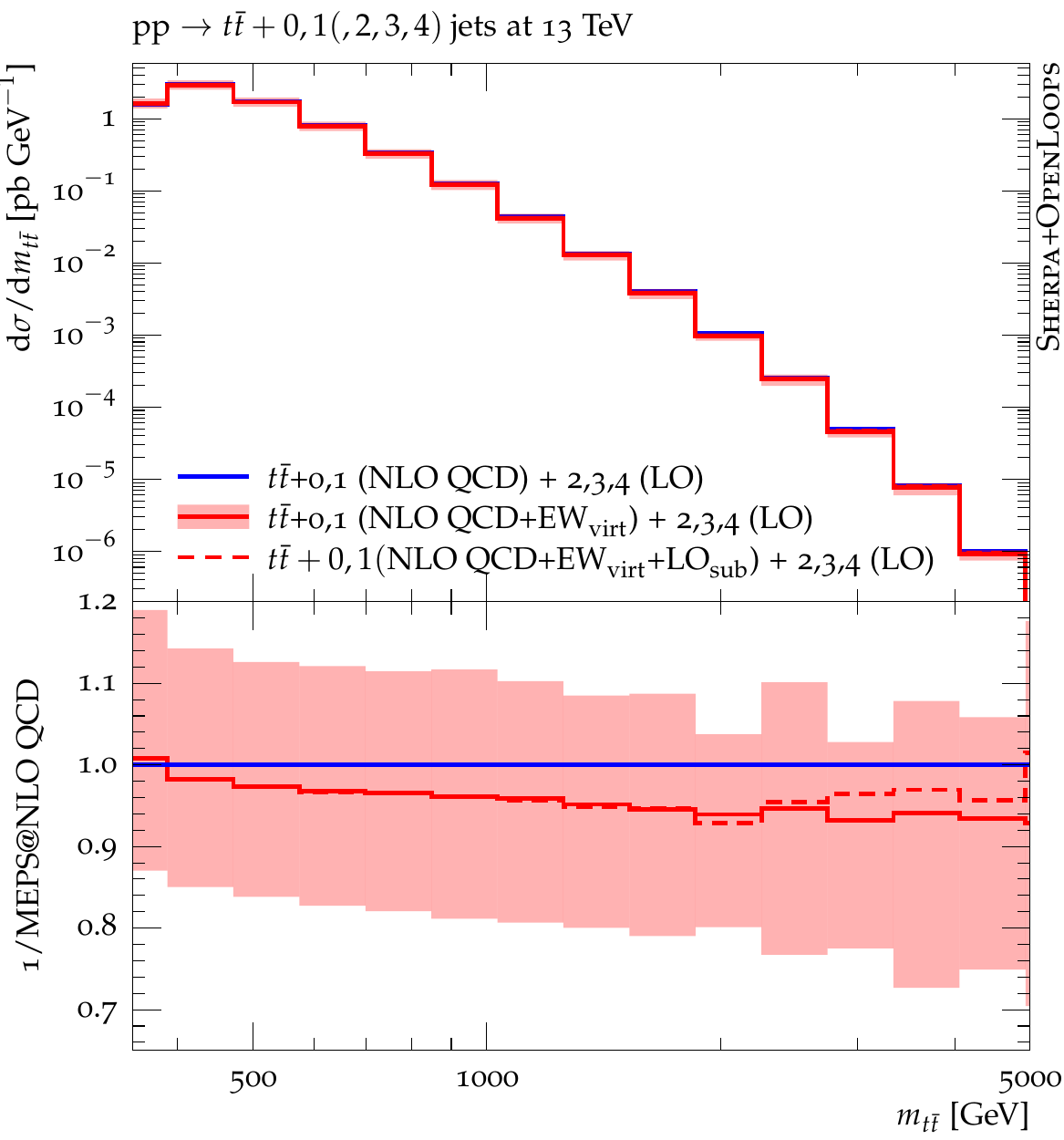}
  \caption{
     Top-quark transverse momentum distribution (left) and top-antitop invariant mass (right) at parton-level 
     with stable tops for the LHC with 13 TeV.
     Compared are \MEPSatNLO \QCD and \MEPSatNLO \QCDpEWvirt predictions and the effect of subleading Born contributions. 
     Error bands are due to QCD scale variations.
  }
\label{fig:pTtop_mtt_merged}
\end{figure*}

\begin{figure}[t]
 \centering
   \includegraphics[width=.48\textwidth]{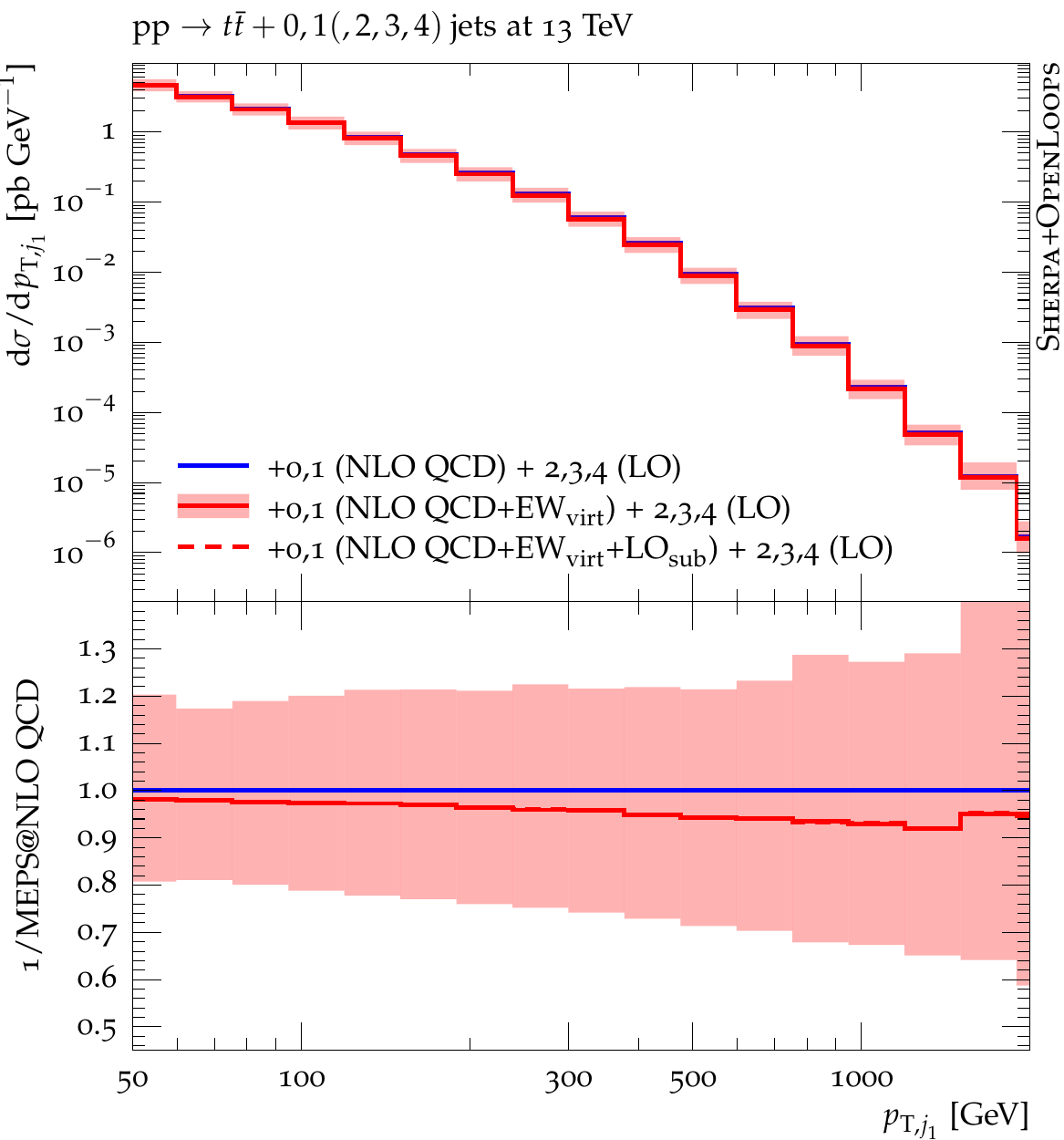}
  \caption{
Leading jet transverse momentum distribution at the LHC with 13 TeV comparing   \MEPSatNLO \QCD
and   \MEPSatNLO \QCDpEWvirt parton-level predictions. Error bands are due to QCD scale variations.
  }
\label{fig:pTjet1_meps}
\end{figure}

In Figures \ref{fig:pTtop_mtt_merged} and \ref{fig:pTjet1_meps} we present parton-level 
multi-jet merged \linebreak \MEPSatNLO \QCDpEWvirt
predictions at the LHC with \linebreak $13~$TeV. Here the top quark is treated as stable and no 
non-perturbative effects are included.
Input parameters and settings are chosen as detailed
in Section~\ref{sec:fo}. We merge $\ttbar$ plus zero and one jet production
based on NLO matrix elements including $\order(\alphaS)$ QCD corrections and $\order(\alpha)$ 
EW corrections in the EW$_{\rm virt}$ approximation. In all merged predictions we choose $Q_{\rm cut}=30$~GeV.~\footnote{
In Appendix B of \cite{Kallweit:2015dum} it was shown that even for the prediction of observables in the multi-TeV
regime the associated uncertainty related to variations of the merging scale are very small.
}
The effect of additionally including the subleading Born contributions of
$\order(\alphaS^{1+n}\alpha)$ and $\order(\alphaS^{n}\alpha^2)$ in the merging
is shown explicitly. In order to allow for a direct comparison with the
corresponding fixed-order results we chose renormalisation and factorisation
sca\-les according to Eq. (\ref{eq:foscale}). The shown error bands indicate
resulting fac\-tor-2 QCD scale variations. In the ratio of the\linebreak
\MEPSatNLO QCD+EW$_{\rm virt}$ predictions over the \MEPSatNLO QCD predictions
we recover EW correction factors consistent with the fixed-order results
presented in Section~\ref{sec:fo}. The same also holds for the effect of the
subleading Born contributions.

\begin{figure}[t]
 \centering
   \includegraphics[width=.48\textwidth]{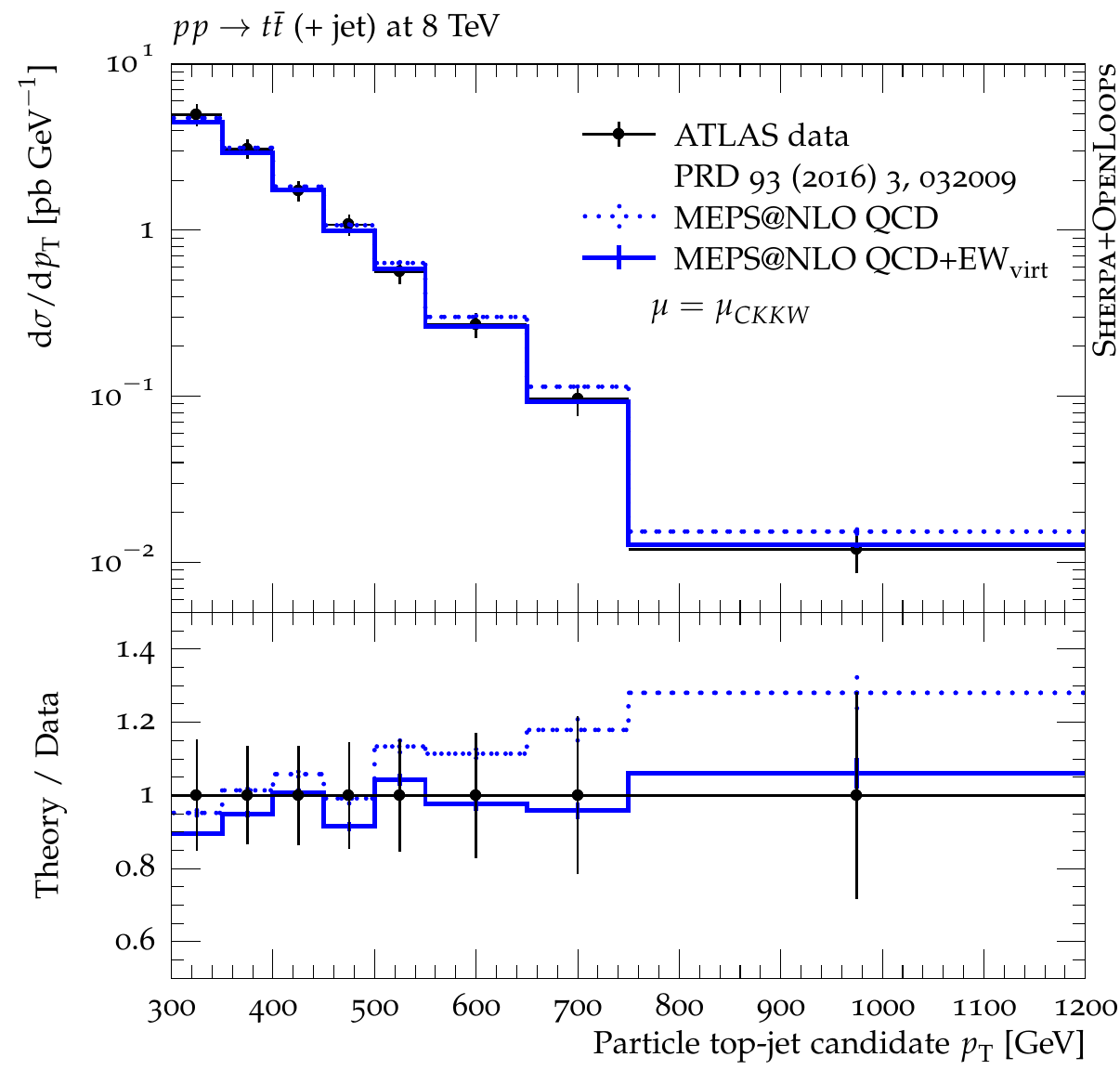}
  \caption{
Comparison of the MEPS@NLO QCD and MEPS@NLO QCD+EW$_{\rm virt}$ predictions for the  transverse momentum distribution of hadronically decaying top candidates
against an 8 TeV ATLAS measurement in the lepton+jets channel based on a boosted top selection \cite{Aad:2015hna} .
  }
\label{fig:pTtop_data}
\end{figure}

Finally, in Figure \ref{fig:pTtop_data} we present full particle-level
\linebreak \MEPSatNLO \QCDpEWvirt predictions for multijet-merged top-pair
production including spin-correlated top quark decays \cite{Hoche:2014kca} in
the semileptonic decay channel. Here, also non-perturbative effects due to
multiple interaction simulation \cite{Alekhin:2005dx}, hadronisation
\cite{Winter:2003tt} and hadron decays, as well as higher-order QED effects
included through the soft-photon resummation of \cite{Schonherr:2008av} have
been included. These predictions are compared to experimental data taken by the
ATLAS experiment \cite{Aad:2015hna} at the LHC at 8\,TeV measuring the
transverse momentum distribution of reconstructed top-quark candidates. The
corresponding analysis is implemented in \Rivet \cite{Buckley:2010ar} and
entails a reconstruction of the transverse momentum of the hadronically decaying
top-quark candidates with $\pT>300$\,GeV.
In this measurement, the boosted top-quark candidate is identified as 
a single large-radius jet ($R=1.0$) using jet substructure techniques.

We find a significant improvement of the agreement between MC simulation
and data when electroweak corrections are included, although the statistical 
prowess of the data sample as well as the high-$\pT$ reach 
are limited in this measurement.

\section{Conclusions}
\label{sec:conclusions}

In this paper we have presented the first predictions for top-pair plus jet
production including Born and one-loop EW corrections. We compared these
corrections with the ones for top-pair production and overall found a universal
behaviour indicating a factorisation of the EW corrections with respect to
additional jet radiation. Subsequently, based on the \MEPSatNLO multijet merging
framework in {\sc Sherpa} combined with \OpenLoops, we derived parton- and
particle-level predictions for inclusive top-pair production including NLO QCD
and EW corrections. The EW corrections are incorporated in an approximation,
based on exact virtual NLO EW contributions combined with integrated-out QED
bremsstrahlung. We showed that this approximation is able to reproduce the full
NLO EW result for $t\bar t$ and $t\bar t+$jet production at the percent level.
Comparing our predictions against a recent measurement for the top-quark
$\pT$-spectrum performed by ATLAS in the lepton+jet channel we find very good
agreement between the \NLO Monte Carlo predictions and data when the EW
corrections are included.

\begin{acknowledgements}
We thank Frank Krauss and Stefano Pozzorini for useful conversations. 
The latter we also thank for valuable comments on the manuscript.
This work has received funding from the European Union's Horizon 2020 
research and innovation programme as part of the Marie Skłodowska-Curie 
Innovative Training Network MCnetITN3 (grant agreement no. 722104).
\end{acknowledgements}

\appendix

\bibliographystyle{JHEP}
\bibliography{refs}

\end{document}